\documentclass[twocolumn]{aastex63}

\usepackage{bm}
\usepackage{color}

\submitjournal{ApJ}

\shorttitle{Elimination of the Blue Loops}
\shortauthors{Mori et al.}

\begin{document}

\title{Elimination of the Blue Loops in the Evolution of Intermediate-mass Stars by the Neutrino Magnetic Moment and Large Extra Dimensions}

\correspondingauthor{Kanji Mori}
\email{kanji.mori@grad.nao.ac.jp}

\author[0000-0003-2595-1657]{Kanji Mori}
\altaffiliation{Research Fellow of Japan Society for the Promotion of Science}
\affiliation{Graduate School of Science, The University of Tokyo, 7-3-1 Hongo, Bunkyo-ku, Tokyo, 113-0033 Japan}
\affiliation{National Astronomical Observatory of Japan, 2-21-1 Osawa, Mitaka, Tokyo 181-8588, Japan}

\author[0000-0002-2999-0111]{A. Baha Balantekin}
\affiliation{Department of Physics, University of Wisconsin-Madison, Madison, Wisconsin 53706 USA}
\affiliation{National Astronomical Observatory of Japan, 2-21-1 Osawa, Mitaka, Tokyo 181-8588, Japan}

\author[0000-0002-8619-359X]{Toshitaka Kajino}
\affiliation{National Astronomical Observatory of Japan, 2-21-1 Osawa, Mitaka, Tokyo 181-8588, Japan}
\affiliation{School of Physics, Beihang University, 37 Xueyuan Road, Haidian-qu, Beijing
100083, China}
\affiliation{Graduate School of Science, The University of Tokyo, 7-3-1 Hongo, Bunkyo-ku, Tokyo, 113-0033 Japan}

\author[0000-0003-2305-9091]{Michael A. Famiano}
\affiliation{National Astronomical Observatory of Japan, 2-21-1 Osawa, Mitaka, Tokyo 181-8588, Japan}
\affiliation{Department of Physics, Western Michigan University, Kalamazoo, Michigan 49008 USA}



\begin{abstract}
For searching  beyond Standard Model physics, stars are laboratories  which complement terrestrial experiments. Massless neutrinos in the Standard Model of particle physics cannot have a magnetic moment, but massive neutrinos have a finite magnetic moment in the minimal extension of the Standard Model. Large extra dimensions are a possible solution of the hierarchy problem. Both of these provide additional energy loss channels in stellar interiors via the electromagnetic interaction and radiation into extra dimensions, respectively, and thus affect stellar evolution.  We perform simulations of stellar evolution with such additional energy losses and find that they eliminate the blue loops in the evolution of intermediate-mass stars. The existence of Cepheid stars can be used to constrain the neutrino magnetic moment and large extra dimensions. 
In order for Cepheids to exist, the neutrino magnetic moment should be smaller than the range  $\sim 2\times10^{-10}$ to 
$4\times10^{-11}\mu_\mathrm{B}$
, where $\mu_\mathrm{B}$ is the Bohr magneton, and  the fundamental scale in the (4+2)-spacetime should be larger than {$\sim2$ to 5 TeV,  depending on the rate of the $^{12}$C$(\alpha, \gamma)^{16}$O reaction. The fundamental scale also has strong dependence on the metallicity}. This value of the  magnetic moment is in the range explored in the reactor experiments, but higher than the limit inferred from globular clusters. Similarly the fundamental scale value we constrain corresponds to a size of the compactified dimensions comparable to those explored in the torsion balance experiments, but is smaller than the limits inferred from collider experiments and low-mass stars.
\end{abstract}



\section{Introduction} \label{sec:intro}

Intermediate-mass stars deviate from the red giant branch and form a loop towards the blue region in the Hertzsprung-Russell (HR) diagram during central helium burning \citep{2012sse..book.....K}. Such a loop is called  a ``blue loop.'' Stars spend considerable time on the blue loop, so many blue giants have been discovered \citep[e.g.][]{2011ApJ...740...48M,2002AJ....123.1433D,1993AJ....105.1956E}. The blue loops can cross the Cepheid instability strip if their endpoint extend to high enough temperature. In that case, the stars on the blue loops are observed as Cepheid variables. 

Stars have been used to explore beyond-standard physics which may be difficult to reach with laboratory experiments \citep{1996slfp.book.....R}. Recently, it was pointed out that the blue loops in the evolution of intermediate-mass stars can be eliminated  if energy loss from axion emission \citep{2013PhRvL.110f1101F} is included in stellar evolution calculations.  Because the blue loops are 
a ubiquitous characteristic of blue giants and Cepheid variables, this is a powerful way to relate new physics to observations. We apply this idea  to the exploration of  non-standard energy losses that originate from the neutrino magnetic moment ($\mu_\nu$; NMM) and large extra dimensions (LEDs). 

In the standard model (SM) of particle physics, neutrinos are assumed to be massless. However, neutrino oscillation observations have revealed that they have mass eigenstates \citep[e.g.][]{1998PhRvL..81.1562F}. The NMM is allowed only for massive neutrinos and the minimally extended SM predicts a small but finite magnetic moment \citep{1982NuPhB.206..359S,1980PhRvL..45..963F}. 

Since the NMM is a key to physics beyond the Standard Model, several  experiments have been performed to find it and determine its magnitude \citep{2018ARNPS..68..313B,2015RvMP...87..531G}. The most recent constraint comes from the GEMMA experiment \citep{2013PPNL...10..139B}, which measures the scattering cross sections of electrons and reactor anti-electron neutrinos. This 
constrains the magnetic moment at $\mu_\nu<2.9\times10^{-11}\mu_\mathrm{B}$ (90\% C.L.).

In addition to the intermediate-mass stars considered here NMMs can also be constrained from low-mass stars. The luminosity of the tip of red giant branch  is sensitive to the energy loss. Theoretical luminosities are compared to the color-magnitude diagram of globular clusters \citep{2015APh....70....1A,2013PhRvL.111w1301V, 2013A&A...558A..12V,1992A&A...264..536R} and a stringent constraint, $\mu_\nu<2.2\times10^{-12}\mu_\mathrm{B}$, is reported \citep{2015APh....70....1A}.

The idea of LEDs is proposed by \cite{1998PhLB..429..263A} to solve the hierarchy problem, i.e. the huge difference between the electroweak scale $\sim$ TeV and the Planck scale $\sim10^{16}$ TeV \citep{2018PhRvD..98c0001T}. The  Planck mass $M_\mathrm{S}$ in the $(4+n)$-dimensional spacetime is related with that $M_\mathrm{P}$ in the 4-dimensional spacetime as \citep{1999PhLB..461...34B}
\begin{eqnarray}
M_\mathrm{P}^2=\Omega_n R^nM_\mathrm{S}^{n+2},\label{led}
\end{eqnarray}
where $R$ is the size of the compactified dimensions and $\Omega_n$ is a numerical factor which depends on the geometry of compactification. For example for a torus $\Omega_n = (2 \pi)^n$.  
 In order for the hierarchy problem to be solved, $M_\mathrm{S}$ should coincide with the electroweak scale. For the $n=1$ model, this requires $R\sim10^{10}$ km, which is clearly excluded by the inverse-square law on the scale of the Solar System. In this study, therefore, we focus on the simplest possible case of $n\ge 2$.

The most direct probes of extra dimensions come from torsion balance experiments \citep{2015CQGra..32c3001M,2009PrPNP..62..102A} which measure gravitation at the sub-millimeter range. The gravitational field between two masses $m_1$ and $m_2$ is often parametrized by the Yukawa potential
\begin{eqnarray}
V(r)=-G\frac{m_1m_2}{r}\left(1+\alpha e^{-r/R}\right).
\end{eqnarray}
The $n=2$ model corresponds to $\alpha=16/3$. The most recent torsion experiments report $R\leq 37\;\mu\mathrm{m}$ \citep{2020PhRvL.124e1301T} and $R<30\;\mu\mathrm{m}$ \citep{2020PhRvL.124j1101L}. For $n=2$ this corresponds to a limit of $M_\mathrm{S} \gtrsim 3$ TeV. For $n=3$ this corresponds to a lower limit on $M_\mathrm{S}$ which is well below the electroweak scale. 

Hadron colliders have also been used to search for gravitons. These cannot be directly detected, so energetic jets are examined
for missing transverse energy.
From this, the value of $M_\mathrm{D}$ is extracted, where $M_\mathrm{D}$ is defined as
\begin{eqnarray}
M_\mathrm{P}^2=R^nM_\mathrm{D}^{n+2}.\label{cms}
\end{eqnarray}
The Compact Muon Solenoid (CMS) experiment at the Large Hadron Collider reports $M_\mathrm{D}>9.9$ TeV for $n=2$ \citep{2018PhRvD..97i2005S}. 
This corresponds to a limit of $M_\mathrm{S} \gtrsim 3.9$ TeV.  The corresponding limits from the ATLAS collaboration are slightly lower \citep{2018JHEP...01..126A}.

A more stringent bound comes from  $\gamma$-ray fluxes from neutron stars \citep{2012JCAP...02..012F,2003PhRvD..67l5008H}. A recent observation by the Fermi Large Area Telescope reports a constraint $R<9.5$ nm for the $n=2$ model \citep{2012JCAP...02..012F}. 


Stellar evolution calculations have shown that the tip of the red giant branch is sensitive to LEDs. \cite{2000PhLB..481..323C} conclude that $M_\mathrm{S}>3$ TeV by comparing stars in globular clusters and theoretical stellar evolution. This value is similar to the experimental bounds coming from collider and torsion experiments. Both types of experiments -- with very different systematic errors --
yield bounds in $M_\mathrm{S}$ that are comparable to those derived from evaluations of
the tip of the red giant branch.

Similarly, bounds on neutrino magnetic moments obtained from arguments of energy-loss in low-mass stars are within an order of magnitude of the experimental bounds. 
Terrestrial experiments looking for extra dimensions, such as the torsion balance experiments, and those looking for neutrino magnetic moments are reaching their limits of exploration. To improve the limits on the inverse square law requires a significant increase in the background-free sensitivity for the torsion balance experiments which will be rather difficult. To improve the limits on the neutrino magnetic moment requires ability to measure an exceedingly small amount of the electron recoil energy. Limits from both kind of terrestrial experiments are subject to very different systematic errors as compared to the limits from low-mass stars. Hence it is desirable to explore if other astronomical testbeds can yield limits subject to different systematic errors.
In this paper we explore  bounds obtained from considerations of evolution of intermediate mass stars in the ''blue loop" epoch as these would be subject to different uncertainties than the low-mass stars.

This paper is organized as follows. Section 2 describes the treatment of the extra energy loss due to NMMs and LEDs in stellar models. Section 3 describes the results of stellar evolution calculations. In Section 4, we summarize and discuss the constraints achieved in this study.
\section{Method} \label{sec:style}
\subsection{Energy Loss by the Neutrino Magnetic Moment}
For a non-zero NMM, the neutrino energy loss increases because of an additional electromagnetic contribution to the neutrino emissivity. Here we consider two processes: plasmon decay ($\gamma\rightarrow \nu\bar{\nu}$) and neutrino pair production ($e^+e^-\rightarrow\nu\bar{\nu}$). The additional energy loss rate due to plasmon decay is given as \citep{2009ApJ...696..608H,1994ApJ...425..222H}
\begin{eqnarray}
\epsilon^\mu_\mathrm{plas}=0.318\left(\frac{\omega_\mathrm{pl}}{10\;\mathrm{keV}}\right)^{-2}\left(\frac{\mu_\nu}{10^{-12}\mu_\mathrm{B}}\right)^2\epsilon_\mathrm{plas},\label{plas}
\end{eqnarray}
where $\epsilon_\mathrm{plas}$ is the standard energy loss \citep{1996ApJS..102..411I} and $\omega_\mathrm{pl}$ is the plasma frequency \citep{1996slfp.book.....R}
\begin{eqnarray}
\omega_\mathrm{pl}=28.7\;\mathrm{eV}\frac{(Y_\mathrm{e}\rho)^\frac{1}{2}}{(1+(1.019\times10^{-6}Y_\mathrm{e}\rho)^\frac{2}{3})^\frac{1}{4}}.
\end{eqnarray}
Here $Y_\mathrm{e}$ is the electron fraction and $\rho$ is the density in units of $\mathrm{g\;cm^{-3}}$. The additional energy loss rate due to pair production is written as \citep{2009ApJ...696..608H}
\begin{eqnarray}
\epsilon^\mu_\mathrm{pair}=1.6\times10^{11}\;\mathrm{erg\;g^{-1}\;s^{-1}}\left(\frac{\mu_\nu}{10^{-10}\mu_\mathrm{B}}\right)^2\frac{e^{-\frac{118.5}{T_8}}}{\rho_4},\label{pair}
\end{eqnarray}
where $T_8=T/(10^8\;\mathrm{K})$ and $\rho_4=\rho/(10^4\;\mathrm{g\; cm^{-3}})$.
\subsection{Energy Loss by Large Extra Dimensions}
A possible existence of compactified extra dimensions results in Kaluza-Klein (KK) modes of gravitons $G_\mathrm{KK}$ with mass $m^2_{\bm{n}}={\bm{n}}^2/R^2$, where $\bm{n}$ is the index for the $n^{th}$ KK modes. The KK gravitons can radiate into extra dimensions and thus work as an additional source of the energy loss, while standard model particles are confined to the 4-dimensional subspace. We consider three processes: photon-photon annihilation ($\gamma\gamma\rightarrow G_\mathrm{KK}$), gravi-Compton-Primakoff scattering ($e^-\gamma\rightarrow e^{-}G_\mathrm{KK}$) and gravi-bremsstrahlung ($e^-(Ze)\rightarrow e^- (Ze)G_\mathrm{KK}$).

The numerical formulae for these processes are given in \cite{2015ApJ...809..141H} and \cite{1999PhLB..461...34B}. The energy loss rates for photon-photon annihilation, gravi-Compton-Primakoff scattering and gravi-bremsstrahlung in the nondegenerate condition are given by
\begin{eqnarray}
\epsilon_{\gamma\gamma}=5.1\times10^{-9}T^9_7\rho^{-1}_6\left(\frac{M_\mathrm{S}c^2}{1\;\mathrm{TeV}}\right)^{-4}\;{\mathrm{erg\;g^{-1}\;s^{-1}}},\\
\epsilon_\mathrm{GCP}=4.5\times10^{-6}T^7_7\left(\frac{M_\mathrm{S}c^2}{1\;\mathrm{TeV}}\right)^{-4}\;{\mathrm{erg\;g^{-1}\;s^{-1}}},\\
\epsilon_\mathrm{GB}=5.8\times10^{-3}\bar{Z}^2_7T^3_7\left(\frac{M_\mathrm{S}c^2}{1\;\mathrm{TeV}}\right)^{-4}\;{\mathrm{erg\;g^{-1}\;s^{-1}}},
\end{eqnarray}
respectively. Here $\bar{Z}_7$ is the mean ion charge relative to nitrogen,  $T_7=T/(10^7\;\mathrm{K})$ and $\rho_6=\rho/(10^6\;\mathrm{g\; cm^{-3}})$.
\subsection{Stellar Model}
We use a one-dimensional stellar evolution code Modules for Experiments in Stellar Astrophysics \citep[\texttt{MESA};][]{MESA1,MESA2,MESA3,MESA4,MESA5} version 10398. The code adopts the equation of state of \citet{Rogers2002} and \citet{Timmes2000} and opacities of \cite{Iglesias1996,Iglesias1993} and \cite{Ferguson2005}. 
Nuclear reaction rates are taken from  NACRE \citep{1999NuPhA.656....3A} with weak rates from \cite{Langanke2000,Oda1994,Fuller1985}.  The prescription for screening is based on \cite{Alastuey1978} and \cite{Itoh1979}.

{The initial composition adopted in our models is based on the solar system abundances. Conventionally, the standard solar metallicity has been $Z=0.02$ \citep{1989GeCoA..53..197A}. However, recent literature shows lower metallicities of $Z=0.0122$ \citep{2005ASPC..336...25A}, 0.0134 \citep{2009ARA&A..47..481A} and 0.0148 \citep{2019arXiv191200844L}. In our models, we adopt two compositions: $(Y,\;Z)=(0.28,\;0.02)$ from \citet{1989GeCoA..53..197A} and $(Y,\;Z)=(0.2463,\;0.0148)$ from \citet{2009ARA&A..47..481A}. We call these models Case A and Case B, respectively (Table \ref{case}).}

\begin{table}[tb]
\begin{tabular}{c|ccc|}
&X&Y&Z\\\hline
Case A&0.70&0.28&0.02\\
Case B&0.7389&0.2463&0.0148
\end{tabular}
\caption{The initial composition adopted in our models. \label{case}}
\end{table}

Convective mixing
lengths are fixed to $\alpha=1.6$, {which were adopted in \citet{2013PhRvL.110f1101F}.}
The overshoot parameter is set to be $f_\mathrm{ov}=0.005$. {When the effective temperature $T_\mathrm{eff}$ is lower than $10^4$ K, the mass loss table compiled by \citet{1988A&AS...72..259D} is used. When $T_\mathrm{eff}$ is higher than $10^4$ K, mass loss is not taken into account. Pulsation-driven mass loss \citep{2011A&A...529L...9N,2008ApJ...684..569N} within the Cepheid instability strip is not considered.} The nuclear reaction network includes 22 nuclides (\texttt{approx21\_plus\_co56.net}). Evolution is followed until the end of core helium burning.

\section{Results}
We calculate {non-rotating} stellar models with 7, 8, 9, and 10$M_\odot$\footnote{{Models heavier than $10M_\odot$ do not undergo the blue loops with the adopted parameters.}}. The adopted NMM is $\mu_{12}=100$, 200 and 300, where
$\mu_{12}$ is the neutrino magnetic moment in units of $10^{-12} \mu_B$
\footnote{As noted in the Introduction these values of the  magnetic moment is in the range explored in the reactor experiments, but higher than the limit inferred from globular clusters.}, and the LED adopted mass scales are $M_\mathrm{S}=3$, 2, and 1 TeV\footnote{These values are smaller than those inferred from collider experiments and low-mass stars, but they correspond to the size of compactified dimension currently explored in the torsion balance experiments.}. 
In Section 3.1, we show the HR diagrams of these models. In Section 3.2, we discuss the evolution of the helium burning core and contribution of each elementary process to the energy loss.
\subsection{Elimination of the Blue Loops}
\subsubsection{Case A}
\begin{figure}
\includegraphics[width=8cm]{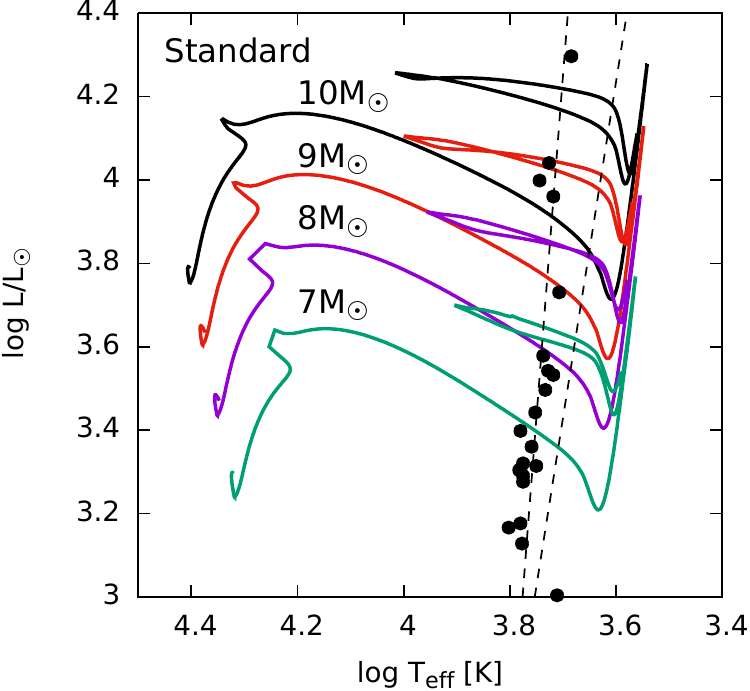}
\includegraphics[width=8cm]{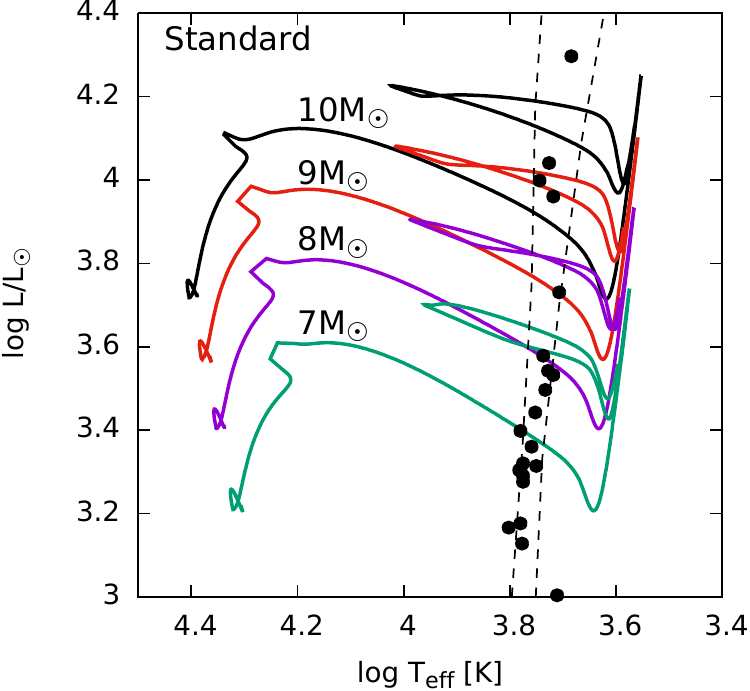}
\caption{The HR diagram with standard physics. The upper panel shows the models with $Z=0.02$ (i.e. Case A) and the lower shows the models with $Z=0.0148$ (i.e. Case B).  The broken lines indicate the edges of the instability strip in models. The $Z=0.02$ models in \citet{2000ApJ...529..293B} are adopted for Case A and the $Z=0.014$ model in \citet{2016A&A...591A...8A} is adopted in  Case B. The points are samples of Galactic Cepheids reported in \citet{2002AJ....124.2931T}.}
\label{fig:hr_std}
\end{figure}

{The upper panel in} Fig. \ref{fig:hr_std} is the HR diagram for the standard case. In this case, all of the models with 7-10$M_\odot$ show the blue loops. The loops in this mass range cross the Cepheid instability strip, in which stars pulsate as Cepheid variables.
\begin{figure*}
\centering
\includegraphics[width=5.5cm]{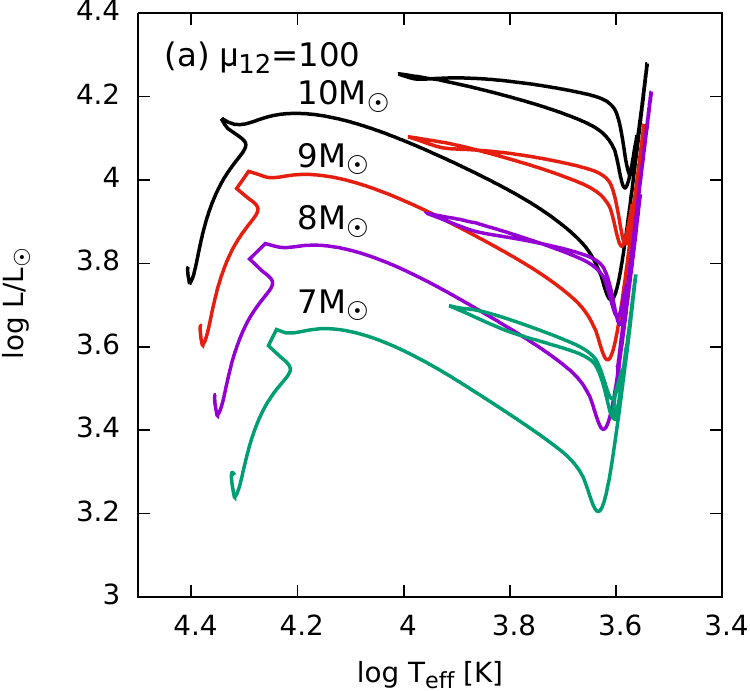}
\includegraphics[width=5.5cm]{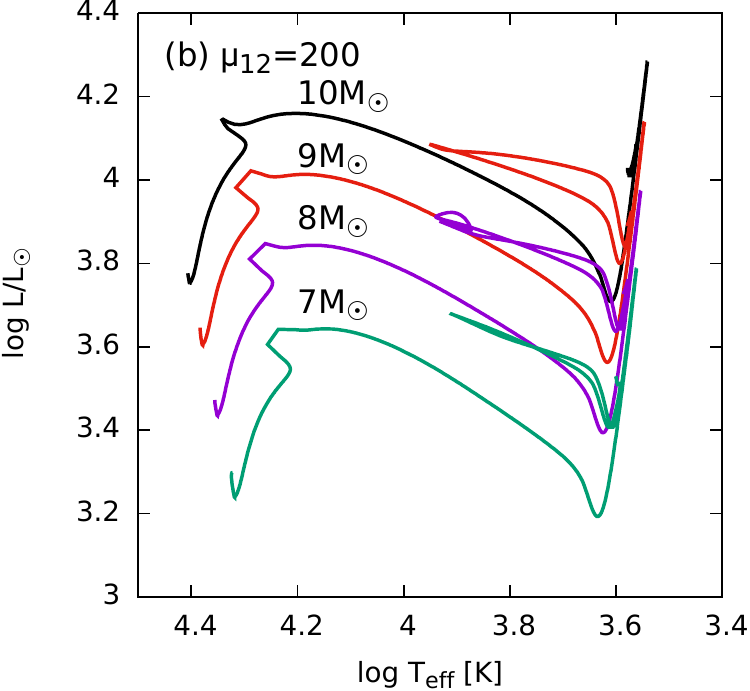}
\includegraphics[width=5.5cm]{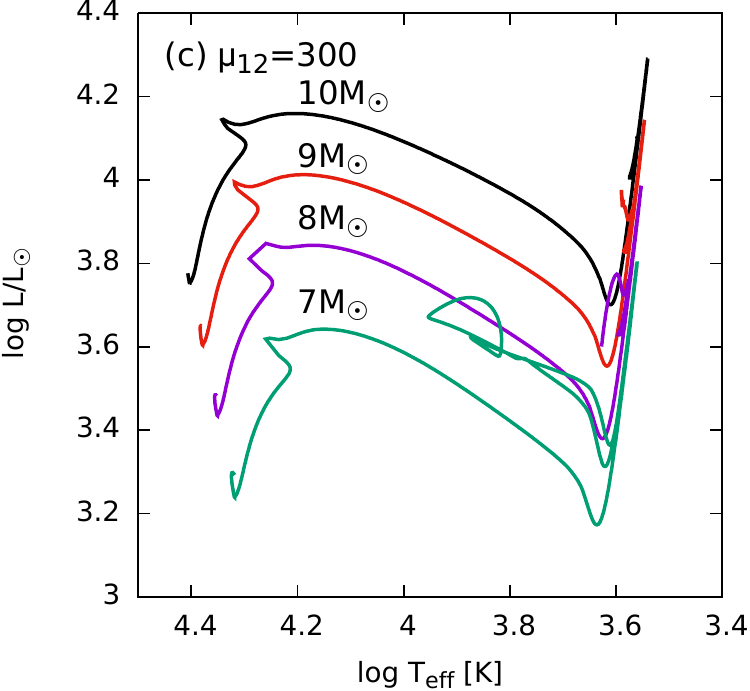}
\caption{The HR diagram with the NMMs of (a) $\mu_{12}=100$, (b) $\mu_{12}=200$ and (c) $\mu_{12}=300$ in Case A.}
\label{fig:hr_mag}
\end{figure*}
\begin{figure*}
\centering
\includegraphics[width=5.5cm]{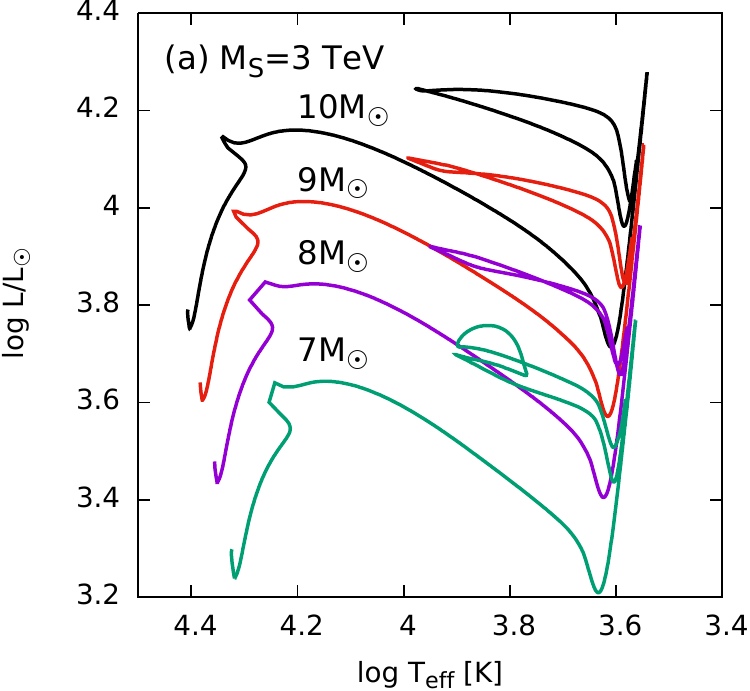}
\includegraphics[width=5.5cm]{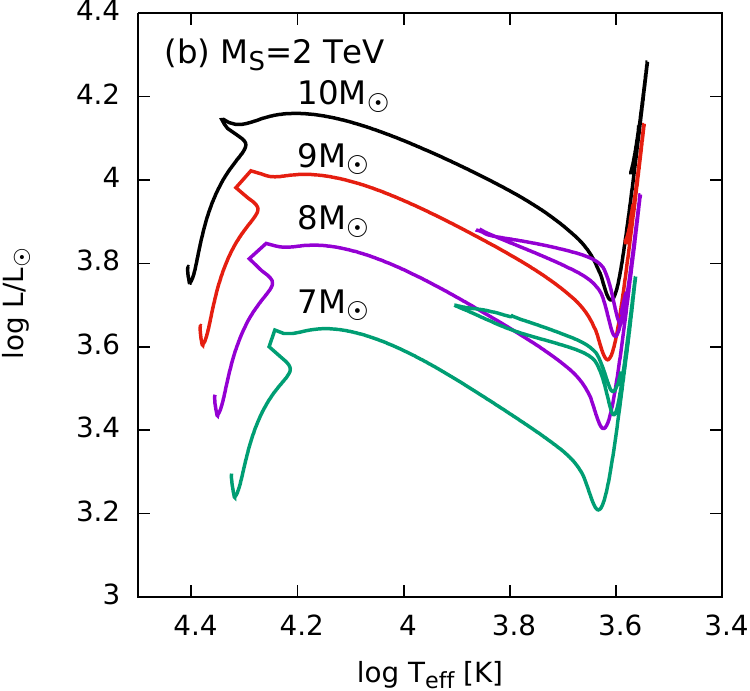}
\includegraphics[width=5.5cm]{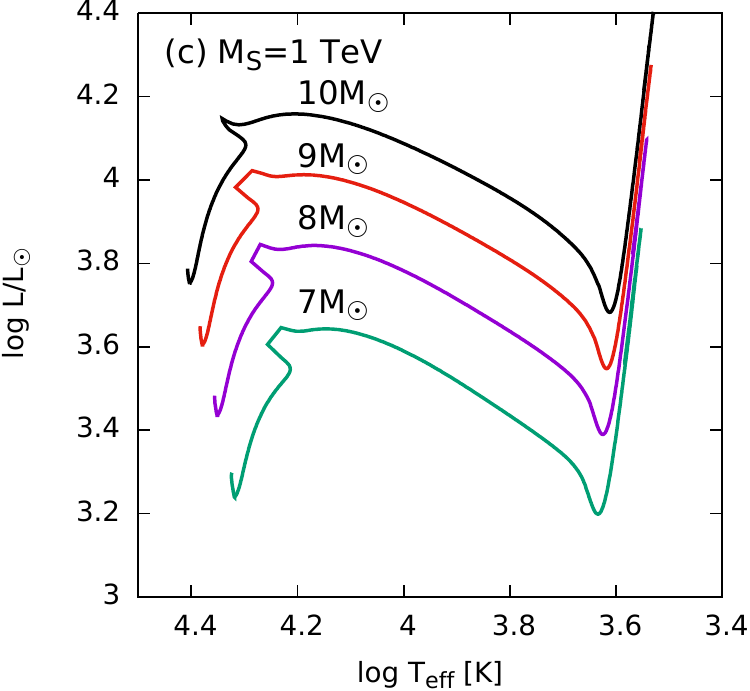}
\caption{The HR diagram with LED of (a) $M_\mathrm{S}=3$ TeV, (b) $M_\mathrm{S}=2$ TeV and (c) $M_\mathrm{S}=1$ TeV in Case A.}
\label{fig:hr_grav}
\end{figure*}

Fig. \ref{fig:hr_mag} is the HR diagram of stars with NMMs of $\mu_{12}=100$, 200 and 300. Though the morphology of the blue loops does not change when $\mu_{12}=100$, in the case of $\mu_{12}=200$, the loop is eliminated for the $10M_\odot$ star. When the NMM is as large as $\mu_{12}=300$, only the $7M_\odot$ model exhibits a blue loop, while its morphology is significantly affected. 

Fig. \ref{fig:hr_grav} shows HR diagrams of stars with LEDs of $M_\mathrm{S}=3$, 2 and 1 TeV. It is seen that, when $M_\mathrm{S}=3$ TeV, the blue loops remain in all of the models, but the morphology is affected for the $7M_\odot$ model. In the case of $M_\mathrm{S}=2$ TeV, the loop is eliminated for the $10M_\odot$ and $9M_\odot$ stars. When $M_\mathrm{S}=1$ TeV, the blue loops are eliminated for all of the models. 

\begin{figure*}
\centering
\includegraphics[width=8.5cm]{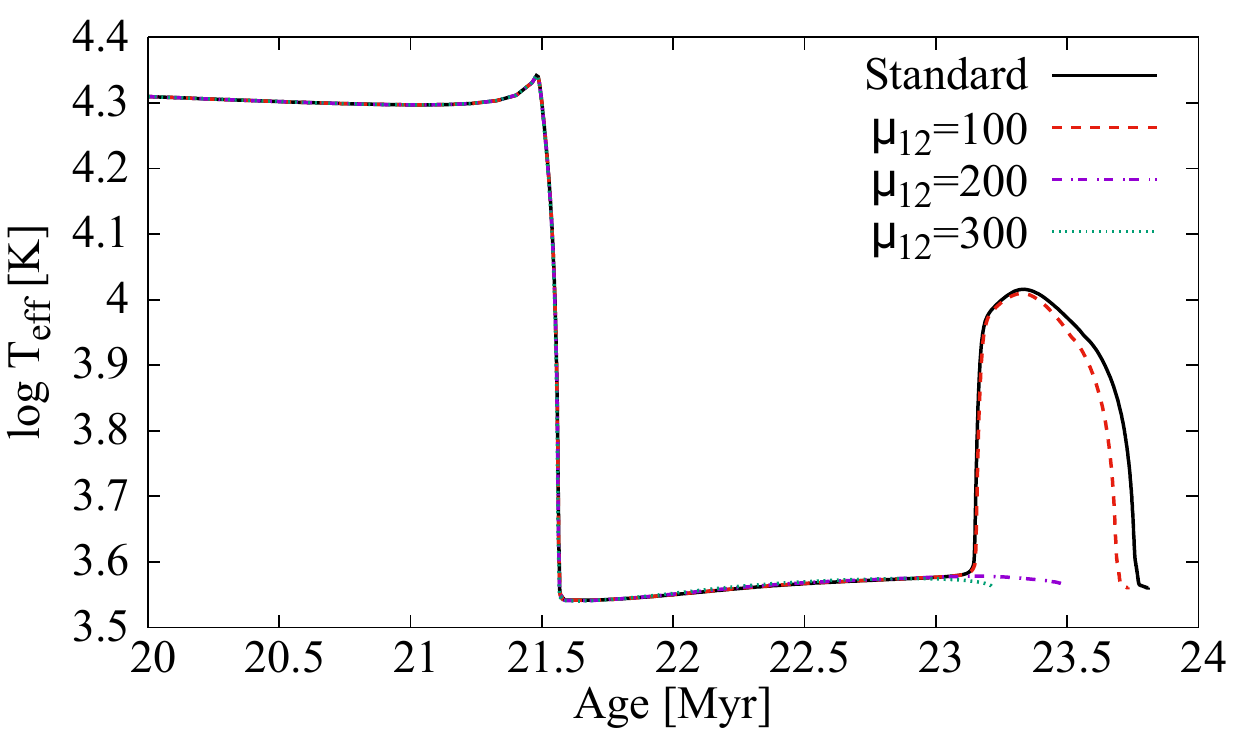}
\includegraphics[width=8.5cm]{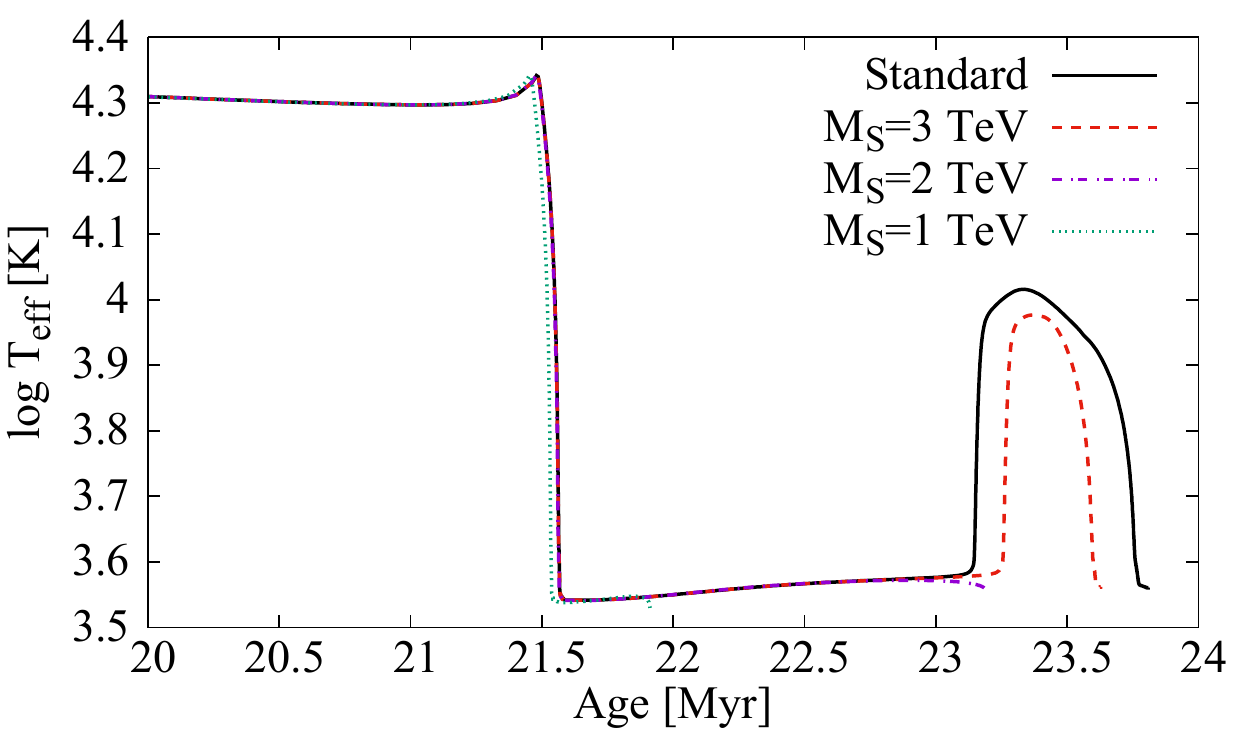}
\caption{The time evolution of the effective temperature {for the $10M_\odot$ models in Case A}. The upper panel shows the effect of the NMM and the lower panel shows the effect of LED.}
\label{fig:teff}
\end{figure*}

Although the blue loops do not disappear for $\mu_{12}=100$ and $M_\mathrm{S}=3$ TeV, the duration, $t_\mathrm{BG}$, of the blue giant phase becomes shorter because of the additional energy loss. Fig. \ref{fig:teff} shows the evolution of the effective temperature as a function of the stellar age for these cases. The upper panel shows the result for various assumptions of the NMM and the lower panel shows the result for various assumptions of LED sizes. The sudden expansion at $\sim21.5$ Myr is the Hertzsprung gap, where the helium core contracts rapidly and the envelope expands \citep{2012sse..book.....K,1952ApJ...116..463S}. The bump around $\sim23.5$ Myr corresponds to the blue loop. It is seen that $t_\mathrm{BG}=0.64$ Myr in the standard case, while $t_\mathrm{BG}=0.55$ Myr when $\mu_{12}=100$ and $t_\mathrm{BG}=0.35$ Myr when $M_\mathrm{S}=3$ TeV. This difference is potentially observable from the ratio of blue and red giants \citep{2011ApJ...740...48M,2002AJ....123.1433D}.
\subsubsection{Case B}
{The HR diagram in the standard case is shown in the lower panel of Fig. \ref{fig:hr_std}. The blue loops appear in all of the models with 7-10$M_{\odot}$. The edges of the blue loops are bluer than those in Case A.}

\begin{figure*}
\centering
\includegraphics[width=5.5cm]{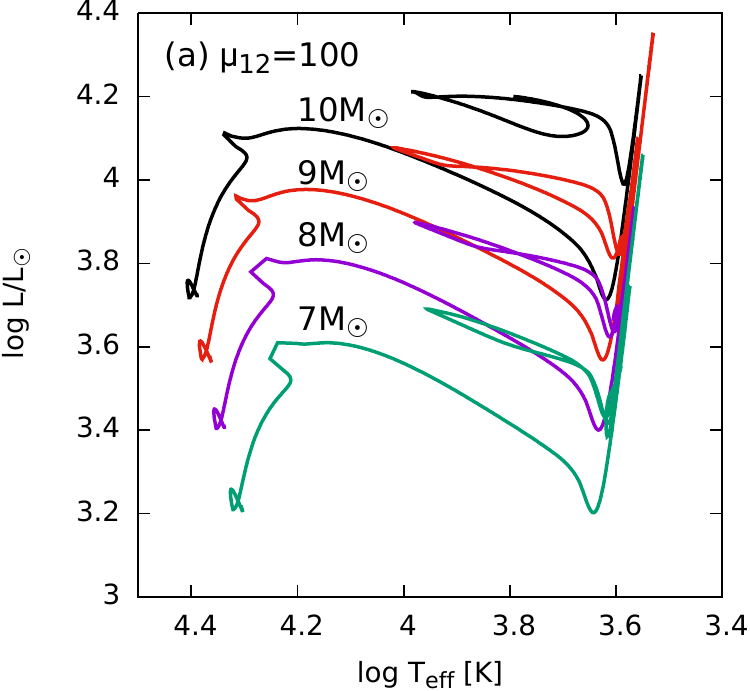}
\includegraphics[width=5.5cm]{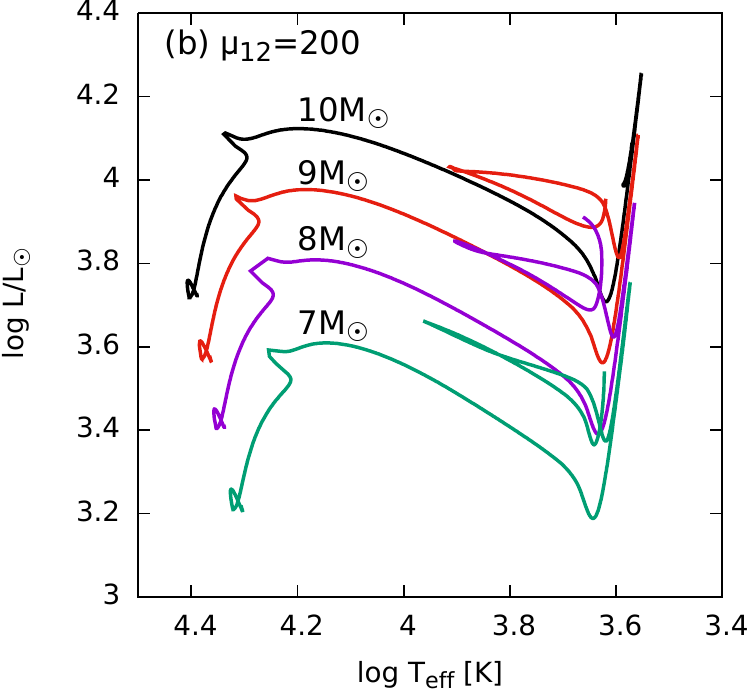}
\includegraphics[width=5.5cm]{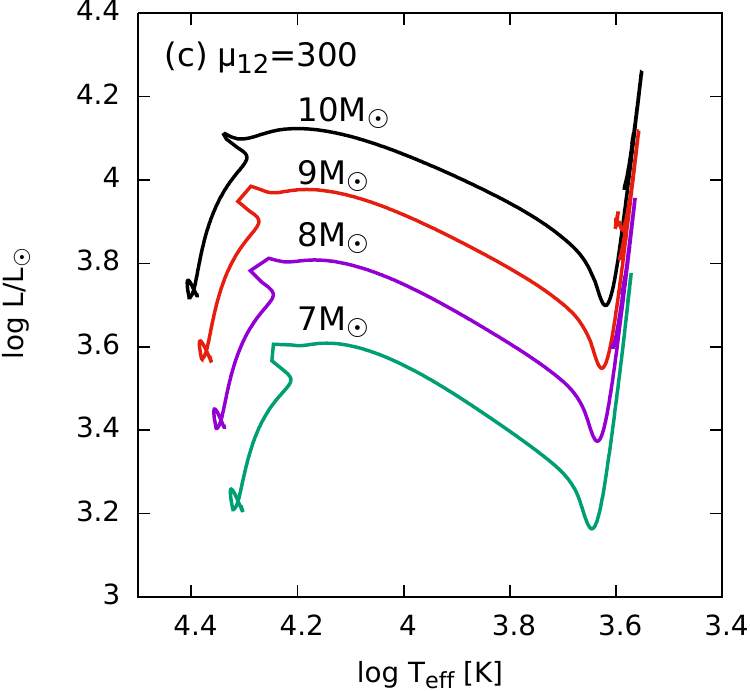}
\caption{The HR diagram with LED of (a) $\mu_{12}=100$, (b) $\mu_{12}=200$ and (c) $\mu_{12}=300$ in Case B.}
\label{fig:hr_mag_solar}
\end{figure*}

{Fig. \ref{fig:hr_mag_solar} is the HR diagram with NMMs of $\mu_{12}=100$, 200 and 300. The blue loops remain in the case of $\mu_{12}=100$, while they are eliminated in the $10M_\odot$ model when $\mu_{12}=200$ and in all of the 7, 8, 9 and $10M_\odot$ models when $\mu_{12}=300$.}

\begin{figure*}
\centering
\includegraphics[width=5.5cm]{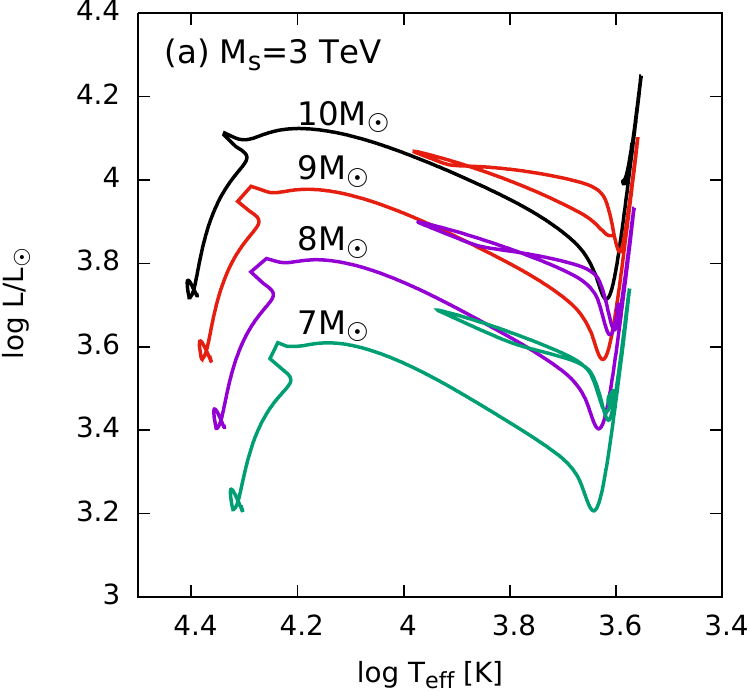}
\includegraphics[width=5.5cm]{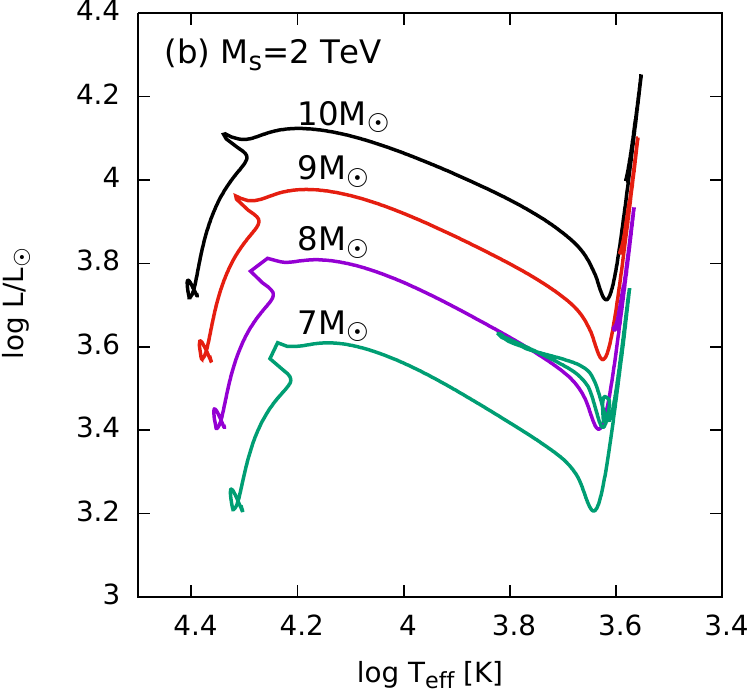}
\includegraphics[width=5.5cm]{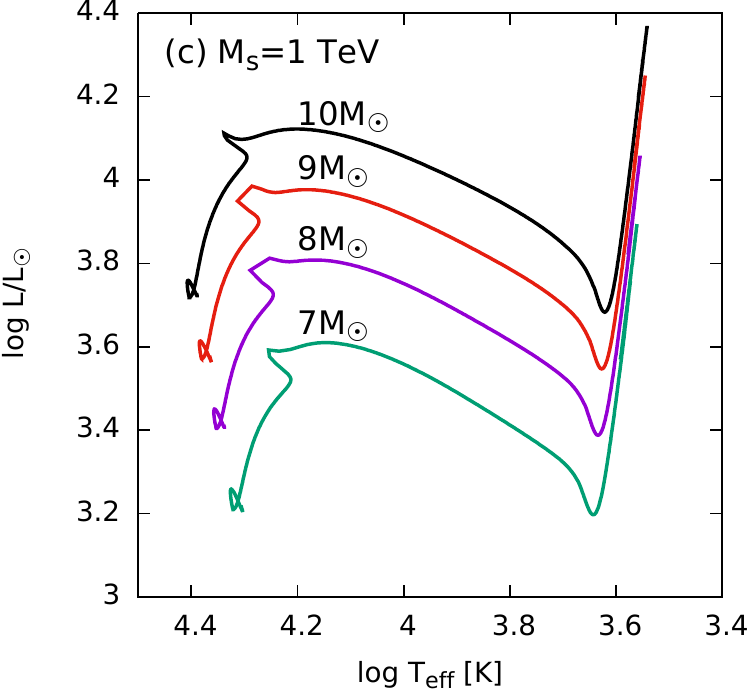}
\caption{The HR diagram with LED of (a) $M_\mathrm{S}=3$ TeV, (b) $M_\mathrm{S}=2$ TeV and (c) $M_\mathrm{S}=1$ TeV in  Case B.}
\label{fig:hr_grav_solar}
\end{figure*}

{Fig. \ref{fig:hr_grav_solar} is the HR diagram with LEDs of $M_\mathrm{S}=3$, 2 and 1 TeV. Contrary to the result in Case A, the blue loop in the $10M_\odot$ model is eliminated even when $M_\mathrm{S}=1$ TeV. The blue loop only in $7M_\odot$ model survives when when $M_\mathrm{S}=2$ TeV and all of the loops are eliminated when $M_\mathrm{S}=3$ TeV. }

\subsection{Evolution of the Core}
\begin{figure}
\centering
\includegraphics[width=8cm]{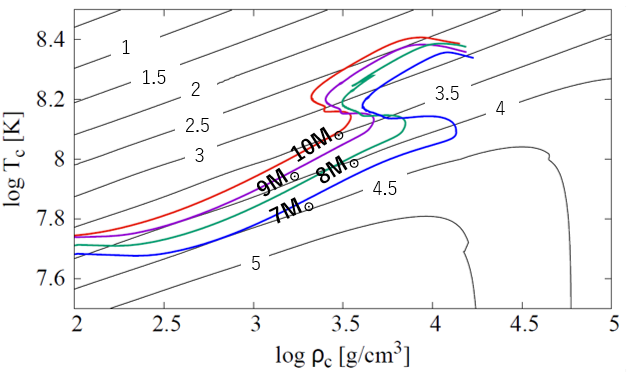}
\includegraphics[width=8cm]{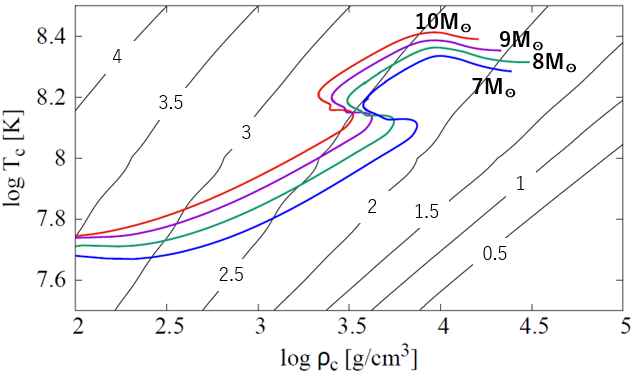}
\caption{The evolution of the central temperature and density in Case A. The upper panel shows the effect of the NMM of $\mu_{12}=200$ and the lower panel shows the effect of LED of $M_\mathrm{S}=2$ TeV. The contour shows the enhancement factor $\log f$ defined in the text.}
\label{fig:center}
\end{figure}

Fig. \ref{fig:center} shows the central temperature and density evolution for stars 
of various mass. The upper panel shows the result with an assumed NMM of $\mu_{12}=200$ and the lower panel shows the result with LED of $M_\mathrm{S}=2$ TeV. The grey contour shows the enhancement factor, $\log f$, of the energy loss defined as
\begin{eqnarray}
\log f=\log\left(\frac{\epsilon_\nu+\epsilon_\mathrm{extra}}{\epsilon_\nu}\right),
\end{eqnarray}
where $\epsilon_\nu$ is the standard energy loss and $\epsilon_\mathrm{extra}$ is the additional energy loss caused by the NMM of $\mu_{12}=200$ and LED of $M_\mathrm{S}=2$ TeV. It is seen that the energy loss rate is enhanced by $10^2$-$10^4$ times.

\begin{figure}
\centering
\includegraphics[width=8cm]{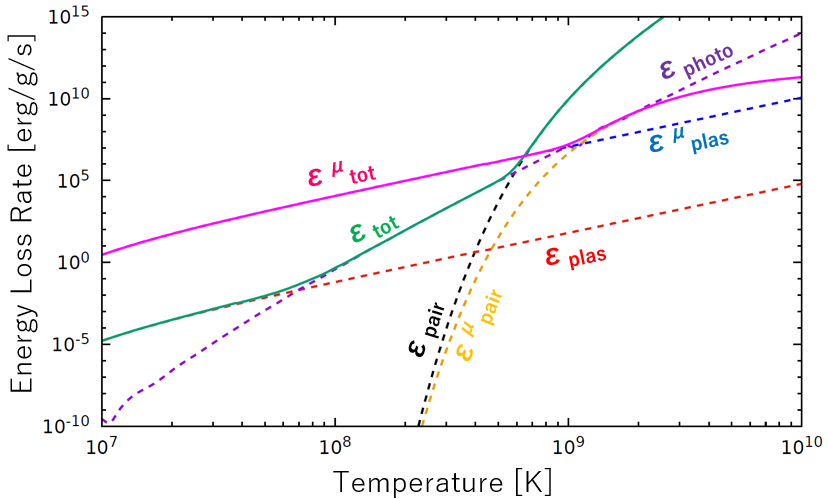}
\includegraphics[width=8cm]{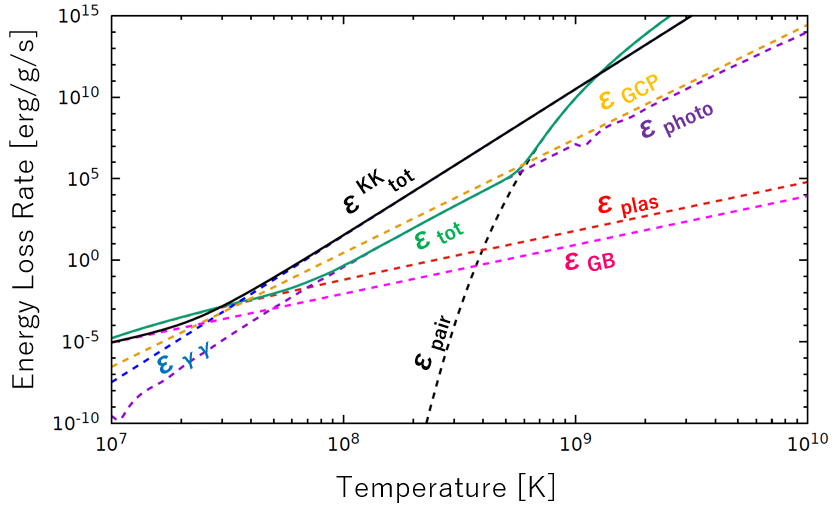}
\caption{Different contributions to the energy loss rates at $\rho=10^4$ g $\mathrm{cm^{-3}}$. The upper panel shows the effect of the NMM of $\mu_{12}=200$ and the lower panel shows the effect of LED of $M_\mathrm{S}=2$ TeV.}
\label{fig:eps}
\end{figure}

From Fig. \ref{fig:center}, one sees that the contribution of $\epsilon_\mathrm{extra}$ decreases as a function of the temperature when $\mu_{12}=200$, while it increases when $M_\mathrm{S}=2$ TeV. This is explained in Fig. \ref{fig:eps}, which shows the energy loss rates of each elementary process at a density of $10^4$ g $\mathrm{cm^{-3}}$.  The upper panel assumes an NMM of $\mu_{12}=200$ and the lower panel assumes a LED of $M_\mathrm{S}=2$ TeV. Here $\epsilon^\mu_\mathrm{plas},\;\epsilon^\mu_\mathrm{pair},\;\epsilon_{\gamma\gamma},\;\epsilon_\mathrm{GCP}$ and $\epsilon_\mathrm{GB}$ are defined in Section 2.
The values of $\epsilon^\mu_\mathrm{tot}$ and $\epsilon^\mathrm{KK}_\mathrm{tot}$ are the total energy loss due to the NMM and LED, respectively. The values $\epsilon_\mathrm{pair},\;\epsilon_\mathrm{plas}$ and $\epsilon_\mathrm{tot}$ are the standard neutrino energy losses \citep{1996ApJS..102..411I}. In the case of $\mu_{12}=200$, the dominant process at $\log T\sim8.2$, where helium burning occurs, is plasmon decay. On the other hand, in the case of $M_\mathrm{S}=2$ TeV, the dominant process is photon-photon annihilation. The photoneutrino energy loss rate $\epsilon_\mathrm{photo}$ is proportional to $T^8$ \citep{1967PhRv..154.1445P}, while the plasma energy loss rate $\epsilon^\mu_\mathrm{plas}$ is proportional to $T^3$ \citep{1964ApJ...140.1025I}. This is the reason why $f$ becomes smaller in the hot region when $\mu_{12}=200$. On the other hand, the photon-photon annihilation rates $\epsilon_{\gamma\gamma}$ is proportional to $T^9$ \citep{1999PhLB..461...34B}, therefore $f$ is larger in the hot region when $M_\mathrm{S}=2$.

The physical mechanism at the onset of the blue loops is still under debate \citep[e.g.][]{2012sse..book.....K,2004A&A...418..213X}. One possible mechanism is the so-called mirror reflection principle. When a star leaves the red giant branch to the blue loop, nuclear burning energy is used to expand the core \citep{2017A&A...605A.106C}. Because of the mirror reflection principle, the expansion of the core leads to the contraction of the envelope and thus higher effective temperature.  However, the NMM and LED extract energy from the core and prevent the expansion of the core. Therefore a star cannot start a detour to a blue giant. 

\begin{figure}
\centering
\includegraphics[width=8cm]{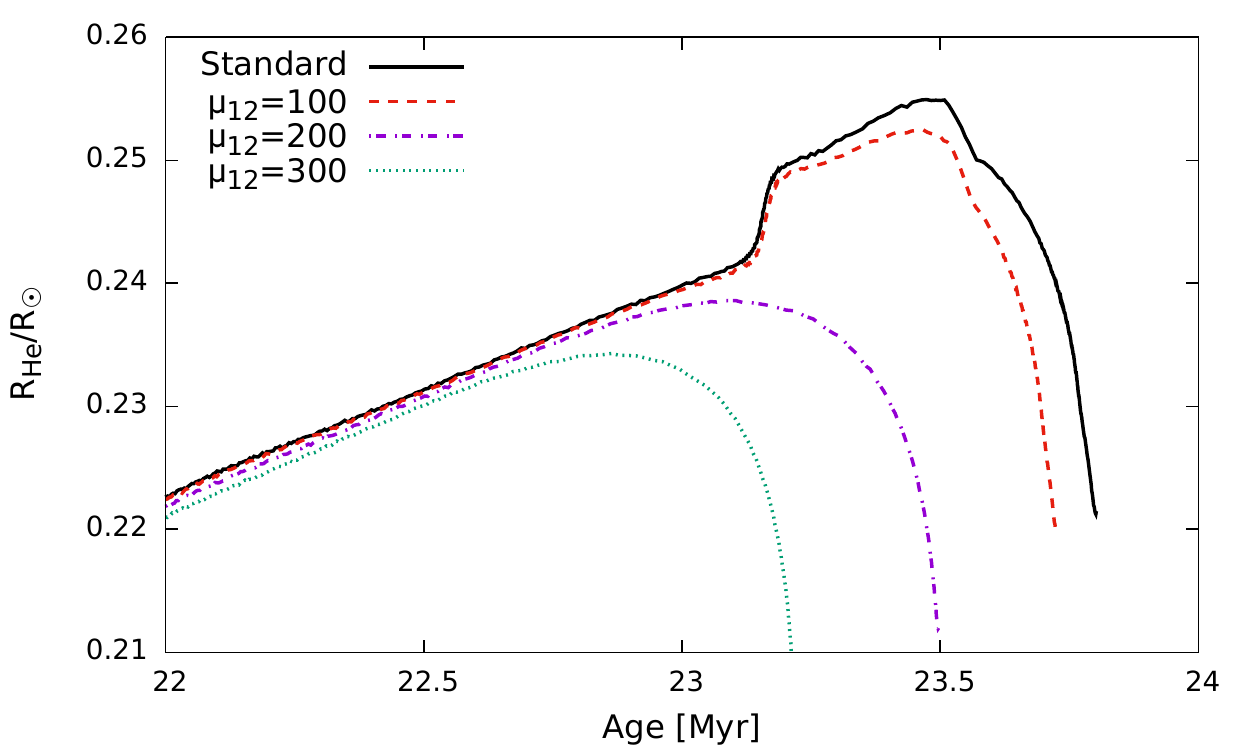}
\caption{The radii of the helium core as a function of stellar age {for the $10M_\odot$ models in Case A}. The solid line shows the result without the NMM and the others show the results for $\mu_{12}=100,\;200$ and 300.}
\label{fig:radius}
\end{figure}

Fig. \ref{fig:radius} shows the evolution of the helium core radius with different NMMs. It is seen that the core radius $R_\mathrm{He}$ increases after $\sim23.2$ Myr in the case of $\mu_{12}=0$ and 100, while it decreases when $\mu_{12}=200$ and 300. This is consistent with the explanation of the blue loop by the mirror reflection principle.

\subsection{Effects of Reaction Rate Uncertainties}
{In the fiducial models, we adopt the NACRE reaction rates \citep{1999NuPhA.656....3A}. However, uncertainties in nuclear reaction rates can significantly affect morphology of the blue loops \citep{2009A&A...507.1541V,2004A&A...418..213X,1990ApJ...351..258B} and thus the threshold of elimination of the loops. In this section, we study the effects of uncertainties in the triple-$\alpha$ and $^{12}$C$(\alpha,\;\gamma)^{16}$O reactions, which govern core helium burning.}
\subsubsection{Triple-$\alpha$ Reaction}
{NACRE estimates temperature-dependent uncertainties in the triple-$\alpha$ reaction to be $\lesssim20\%$ at $\sim10^8$ K. We adopt this uncertainties to study the sensitivity of the blue loops. }

\begin{figure}
\centering
\includegraphics[width=8cm]{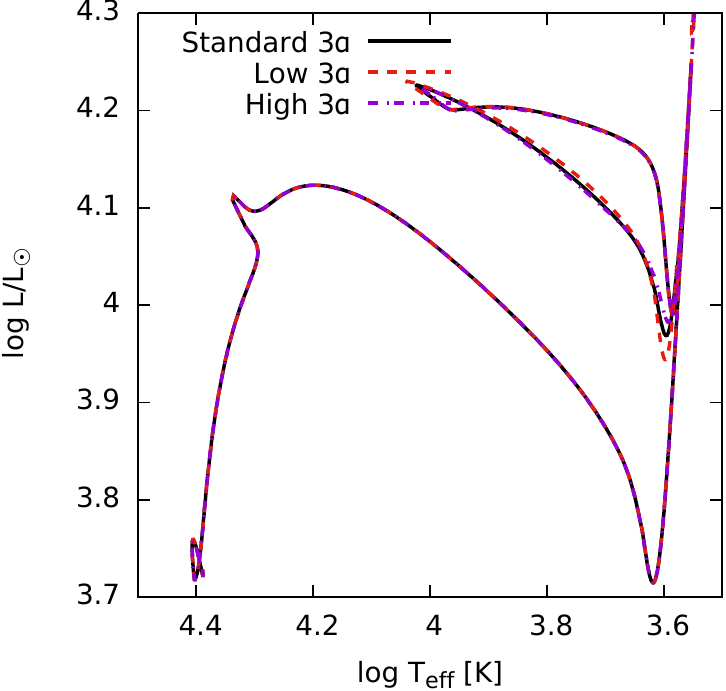}
\caption{The HR diagram of the $10M_\odot$ model with different triple-$\alpha$ reaction rates. The solid line adopts the NACRE standard reaction rate, while the broken lines adopt the higher and lower limits of the rate quoted in NACRE table. The initial composition is set to be Case B.}
\label{fig:HR_3a}
\end{figure}

{Fig. \ref{fig:HR_3a} shows the evolution of the $10M_\sun$ star with the triple-$\alpha$ reactions changed within the NACRE uncertainties. Although the loop extends to the slightly bluer region when the lower rate is adopted, morphology of the blue loops is not affected significantly by the different triple-$\alpha$ rates. This suggests that the threshold of elimination of the loops is robust against  the present uncertainties.}
\begin{figure}
\centering
\includegraphics[width=8cm]{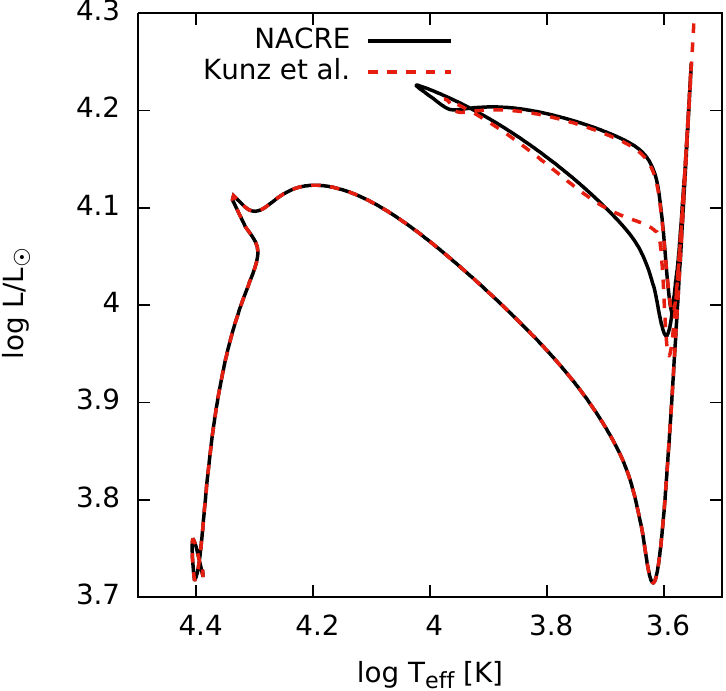}
\caption{The HR diagram of the $10M_\odot$ model with different $^{12}$C$(\alpha,\;\gamma)^{16}$O reaction rates. The solid line adopts the NACRE standard reaction rate, while the broken line adopts the rate quoted in \citet{2002ApJ...567..643K}. The initial composition is set to be Case B.}
\label{fig:HR_kunz2}
\end{figure}
\subsubsection{$^{12}$C$(\alpha,\;\gamma)^{16}$O Reaction}
{The low-energy cross sections of the $^{12}$C$(\alpha,\;\gamma)^{16}$O reaction have not been measured yet \citep[e.g.][]{2017RvMP...89c5007D}. \citet{2002ApJ...567..643K} proposed lower reaction rates than NACRE compilation, based on their new measurements of E1- and E2-capture cross sections. Their reaction rates are $\sim30\%$ smaller than the NACRE rate at $\sim10^8$ K. We adopt the rate recommended by \citet{2002ApJ...567..643K} to perform a sensitivity study.}

{Fig. \ref{fig:HR_kunz2} shows the evolution of the $10M_\odot$ model with the different $^{12}$C$(\alpha,\;\gamma)^{16}$O reaction rates. It is seen that the tip of the blue loop becomes redder when the \citet{2002ApJ...567..643K} rate is adopted and the shape of the loops is significantly different around $\log T_\mathrm{eff}\sim3.65$ between the two.}
\begin{figure*}
\centering
\includegraphics[width=5.5cm]{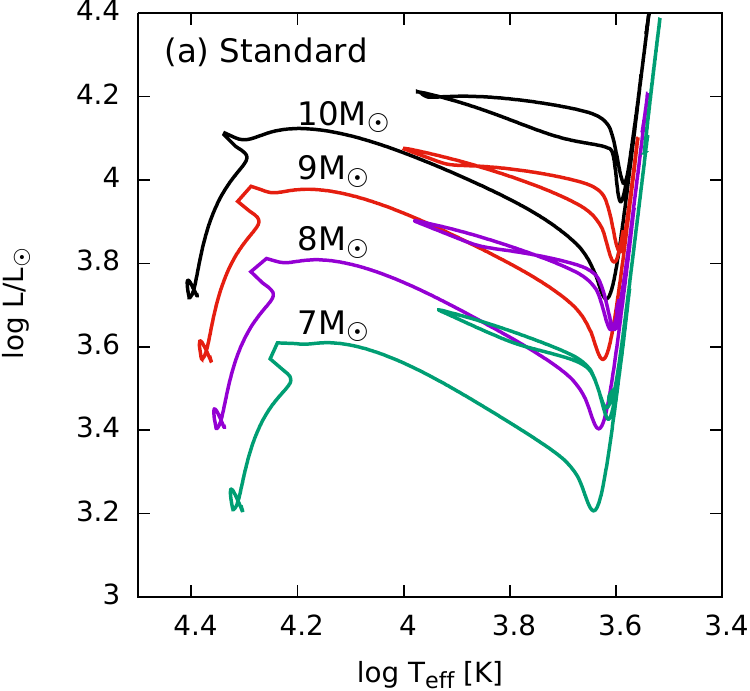}
\includegraphics[width=5.5cm]{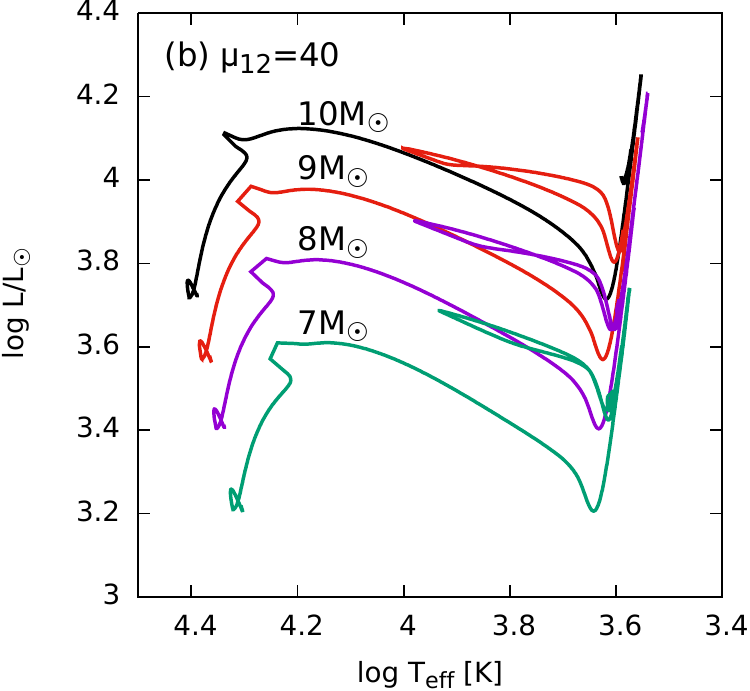}
\includegraphics[width=5.5cm]{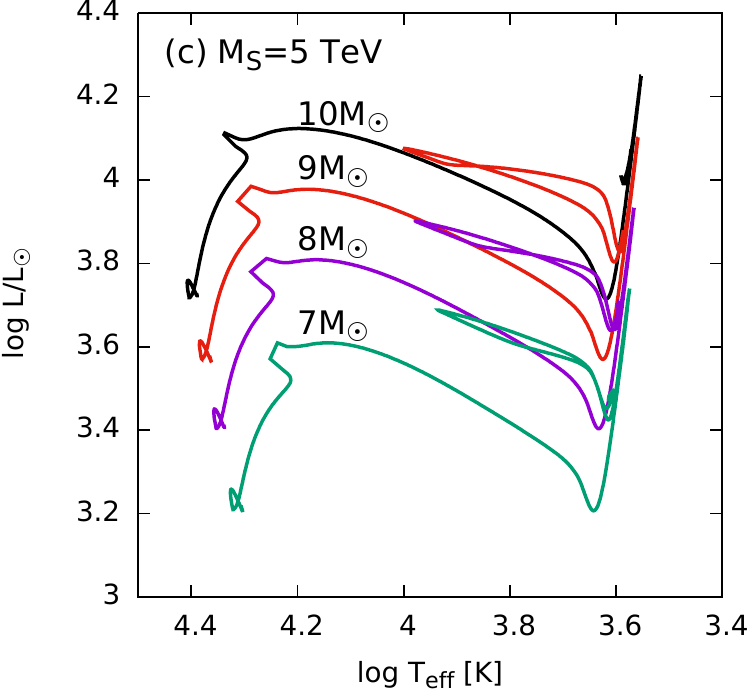}
\caption{The HR diagram (a) without beyond Standard Model physics, (b) with $\mu_{12}=40$, and (c) with $M_\mathrm{S}=5$ TeV. The $^{12}$C$(\alpha,\;\gamma)^{16}$O rate is from \citet{2002ApJ...567..643K}. The initial composition is set to be Case B.}
\label{fig:hr_kunz}
\end{figure*}

{Fig. \ref{fig:hr_kunz} shows the evolution of the 7-10$M_\odot$ stars with the $^{12}$C$(\alpha,\;\gamma)^{16}$O rate quoted in \citet{2002ApJ...567..643K}. When beyond-standard physics is not adopted, the tip of the blue loops becomes redder when  reaction rate is lower, as reported in \citet{2009A&A...507.1541V} and \citet{1990ApJ...351..258B}. Interestingly, the threshold of elimination of the loops is much lower than that with the NACRE rate. As shown in Fig. \ref{fig:hr_kunz}, 
the blue loops are suppressed in the $10M_\odot$ model
 when $\mu_{12}>40$ or $M_\mathrm{S}<5$ TeV with the rate in \citet{2002ApJ...567..643K}, while these thresholds are $\mu_{12}>200$ and  $M_\mathrm{S}<3$ TeV for the NACRE rate, respectively, as has already been discussed in Figs. \ref{fig:hr_mag} - \ref{fig:hr_grav_solar}.}
\subsection{Effects on Heavier Cepheids}
{Some of Galactic Cepheid progenitors have been estimated \citep{1996JRASC..90...82T} to be as massive as $\sim20M_\odot$, using an empirical mass-period relation of Cepheids. Models of such a massive star do not undergo the blue loop during central helium burning \citep[e.g.][]{2016A&A...591A...8A,2009A&A...507.1541V,2000ApJ...543..955B,1992A&AS...96..269S}. Less massive stars with $<15M_\odot$ cross the Hertzsprung gap so rapidly that there is little chance to observe those in the instability strip. However, massive stars with $>15M_\odot$ achieve the central temperature high enough to ignite helium burning before they reach the red giant branch. In this case, the time to cross the gap slows down, so it becomes more probable to observe them in the instability strip.}

\begin{figure}
\centering
\includegraphics[width=8cm]{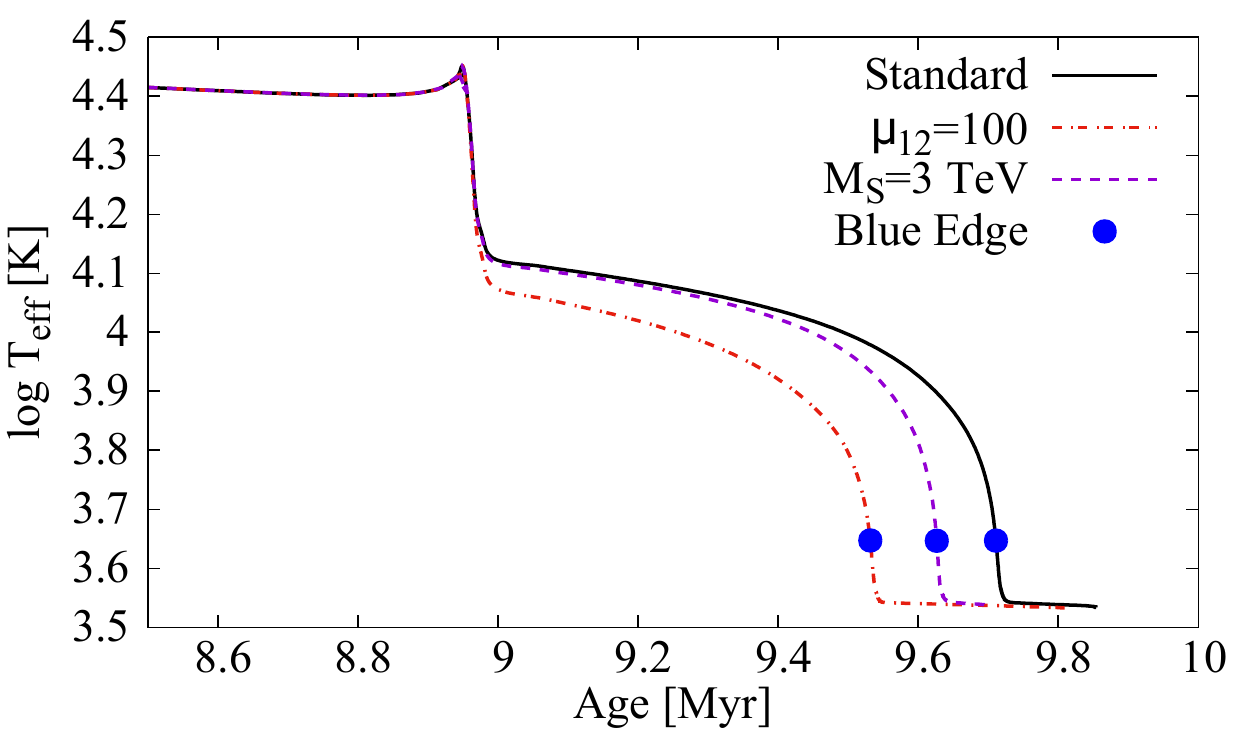}
\caption{Time evolution of the effective temperature for the $20M_\odot$ models in Case B. The black line shows the standard model, and the purple and red lines adopt $\mu_{12}=100$ and $M_\mathrm{S}=3$ TeV, respectively. The dots represent crossing of the blue edge of the instability strip \citep{2000ApJ...529..293B}.}
\label{fig:teff_20}
\end{figure}

{Fig. \ref{fig:teff_20} shows evolution of the effective temperature for the $20M_\odot$ models in Case B. The black line shows the standard evolution and the purple and red lines adopt $\mu_{12}=100$ and $M_\mathrm{S}=3$ TeV, respectively. It is seen that the extra energy losses shorten the timescale of helium burning. The dots show crossing of the blue edge of the instability strip. Although validity of the extrapolation of the model edge \citep{2000ApJ...529..293B} to higher luminosities is uncertain, the stars spend 10-20 kyr in the instability strip even when $\mu_{12}=100$ or $M_\mathrm{S}=3$ TeV is adopted. Therefore these effects do not contradict the observed rare massive Cepheids.}
\subsection{Possible Effects of Mass Loss and Rotation}
{The purpose of this paper is to show fiducial models of intermediate-mass stars with physics beyond the SM, so exhaustive evaluation of theoretical uncertainties is out of its scope. However, the evolution of intermediate-mass stars is sensitive to other physical processes, which is the very reason why they can potentially used as a probe of new physics.}

\begin{figure}
\includegraphics[width=8cm]{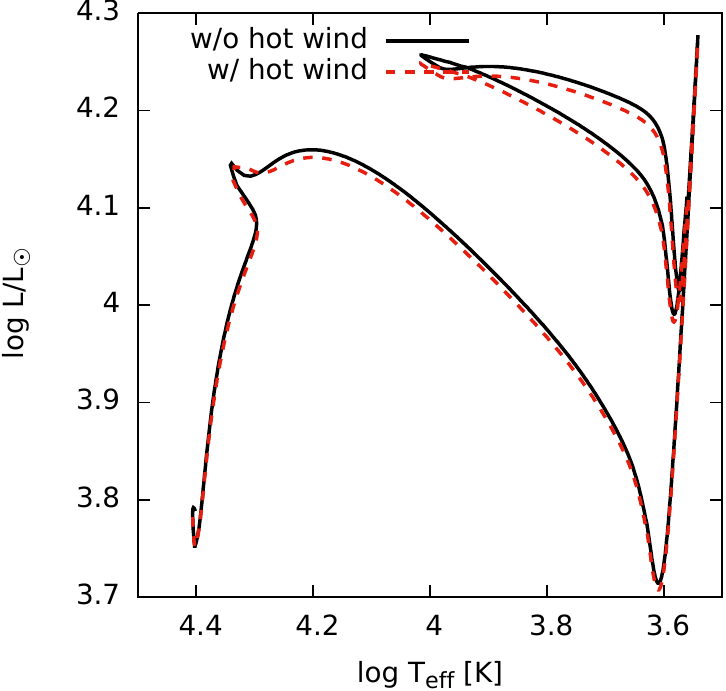}
\caption{The HR diagram for the $10M_\odot$ models with standard physics in Case A. The solid line shows the model without mass loss on the MS, while the broken line shows the model with it.}
\label{fig:hr_massloss}
\end{figure}
{In our models, the treatment of mass loss in the red supergiant phase and the blue loop is based on \citet{1988A&AS...72..259D}, which covers the temperature and the luminosity ranges we are interested in. This mass loss rate on the main sequence (MS) is not considered because it is as small as $10^{-8}$-$10^{-9}M_\odot$ /yr \citep{1988A&AS...72..259D}. Figure \ref{fig:hr_massloss} shows the comparison between the $10M_\odot$ models with and without mass loss during the MS. The solid line shows the fiducial model which was also shown in Fig. \ref{fig:hr_std} and the broken line shows the model with the additional mass loss based on \citet{1988A&AS...72..259D}. It is seen that the mass loss during the MS slightly decreases the luminosity of the blue loop. The effect of the mass loss on the MS on morphology of the blue loops is moderate and thus is not a major source of uncertainties.}

{It has been pointed out that shocks generated by the pulsation drive mass loss up to $10^{-7}M_\odot$ /yr \citep{2008ApJ...684..569N}. Because of such pulsation-driven mass loss, Cepheid variables can lose 5-10 \% of their mass \citep{2011A&A...529L...9N}. Morphology of the blue loops can be significantly affected by  pulsation-driven mass loss, so it is desirable to study uncertainties that originate from it.}

{Effects of rotation are not included in our models. However, the typical rotational velocity of B-stars on the MS with 5-9$M_\odot$ is as high as 10-30 \% of the critical velocity \citep{2010ApJ...722..605H}, so it is important to study the rotational effect. Rotation makes the blue loops more luminous systematically and affect the mass-luminosity relation of Cepheids \citep{2016A&A...591A...8A,2012A&A...537A.146E}. }
\section{Conclusions}
In this paper, we studied the effect of the NMM and LED on the evolution of intermediate-mass stars. We find that the blue loops are eliminated unless $\mu_{12}< 200$ or $M_\mathrm{S}>2$ TeV, placing observational limits on $\mu_{12}$ and $M_\mathrm{S}$. In our models, $10M_\odot$ stars are the most sensitive to beyond-standard physics. 

From Fig. \ref{fig:hr_std}, it is seen that the luminosity of 10$M_\odot$ Cepheids is $\log L/L_\odot\sim4.2$. The period-luminosity relation of Cepheids is written as \citep{1980tsp..book.....C}
\begin{eqnarray}
\log \left(\frac{L}{L_\odot}\right)=1.15\log\left(\frac{P}{1\;\mathrm{day}}\right)+2.47,
\end{eqnarray}
where $P$ is the pulsation period. Putting $\log L/L_\odot\sim4.2$ into this formula, we get $P\sim32$ days. Cepheids with this period are observed in the Galaxy \citep{2006ARA&A..44...93S,2000A&AS..143..211B,1996JRASC..90...82T}.  The existence of $10M_\odot$ Cepheids places an independent
constraint on the NMM and LEDs. 

The current constraints that come from ground experiments are $\mu_{12}<29$ \citep{2013PPNL...10..139B}. Depending on the $^{12}$C$(\alpha,\gamma)^{16}$O rate our constraint on the NMM is either somewhat weaker or comparable to the experimental limit, but higher than the limit inferred from globular clusters.

Using Eq. (\ref{led}), the lower limit on $M_\mathrm{S}$ is transformed to an upper limit  {$R< 30$}  to $170$ $\mu$m compared to the result $R<30\;\mathrm{\mu m}$ reported by the torsion experiment \citep{2020PhRvL.124j1101L}. Eq. (\ref{cms}) shows that the constraint $M_\mathrm{D}>9.9$ TeV, which was reported by the CMS experiment \citep{2018PhRvD..97i2005S}, is equivalent to an upper limit of $R<24\;\mathrm{\mu m}$. {This is to be compared with our result of $M_\mathrm{S} >2$ to $5$ TeV.} The fundamental scale value we constrain corresponds to the size of the compactified dimensions comparable to those explored in the torsion balance experiments, but is smaller than the limits inferred from collider experiments and low-mass stars. In the above results the range depends on the input values of both $^{12}$C$(\alpha,\gamma)^{16}$O rate and the metallicity.

In this study, we focused on the $n=2$ case. We also performed calculations with $n=3$ extra dimensions, using formulae shown in \citet{2000PhLB..481..323C} and \citet{1999PhLB..461...34B}. It is found that the blue loop of a $10M_\odot$ star is eliminated when $M_\mathrm{S}\leq 60$ GeV. Therefore the mass scale for the $n=3$ case can be constrained to be $M_\mathrm{S}>60$ GeV. The CMS experiment \citep{2018PhRvD..97i2005S}, on the other hand, reports $M_\mathrm{D}>7.5$ TeV, so collider experiments can achieve much tighter constraints than energy-loss arguments do in the $n=3$ case.

More quantitative constraints could be achieved by arguments on the timescale of stellar evolution. We saw that the duration of blue giants is shorter when the NMM or LED is included. In order to compare the results with observations, it is desirable to draw isochrones and to superpose them on the color-magnitude diagram. To do so, one must perform calculations with finer grids of stellar masses. The quantitative approach can potentially tighten the constraints on the NMM and LED, but this is beyond the scope of this paper.  

The morphology of the blue loops is very sensitive to input physics including nuclear reaction rates and treatment of metallicity \citep{2012ApJ...761...10H,2011ApJ...741...61S,2010A&A...520A..41M,2009A&A...507.1541V,2004A&A...418..213X,2004A&A...418..225X}.Our results show that there are theoretical uncertainties which originate from these ingredients. To tighten the bounds we obtained it is desirable to perform systematic studies on the effects of different input physics on the constraints of beyond-standard physics. 

\acknowledgments

The authors would like to thank Wako Aoki and Hirokazu Sasaki for fruitful discussions. K.M. is supported 
by JSPS 
KAKENHI Grant Number JP19J12892. 
T.K. is supported in part by Grants-in-Aid for Scientific Research of JSPS (17K05459, 20K03958). M.K. is supported by NSFC Research Fund for International Young Scientists (11850410441). A.B.B. is supported in part by the U.S. National Science Foundation Grant No. PHY-1806368.  M.A.F. is supported by National Science Foundation Grant No. PHY-1712832 and by NASA Grant No. 80NSSC20K0498.  M.A.F. and A.B.B. acknowledge support from the NAOJ Visiting Professor program.
%



\software{MESA \citep{MESA1,MESA2,MESA3,MESA4,MESA5}
          }






\end{document}